\documentclass[conference]{IEEEtran}
\IEEEoverridecommandlockouts
\usepackage{cite}
\usepackage{comment}
\usepackage{amsmath,amssymb,amsfonts}
\usepackage{algorithmic}
\usepackage{graphicx}
\usepackage{textcomp}
\usepackage{xcolor}
\def\BibTeX{{\rm B\kern-.05em{\sc i\kern-.025em b}\kern-.08em
    T\kern-.1667em\lower.7ex\hbox{E}\kern-.125emX}}
\begin{document}

\title{Checkpoint-Restart Libraries Must Become More Fault Tolerant \\

}

\author{\IEEEauthorblockN{Anthony Skjellum}
\IEEEauthorblockA{\textit{University of Tennessee at Chattanooga} \\
Tennessee, USA \\
Tony-Skjellum@utc.edu}
\and
\IEEEauthorblockN{Derek Schafer}
\IEEEauthorblockA{\textit{University of Tennessee at Chattanooga} \\
Tennessee, USA \\
derek-schafer@utc.edu}
}

\maketitle

\begin{abstract}
Production MPI codes need checkpoint-restart (CPR) support.
Clearly, checkpoint-restart libraries  must be fault tolerant lest they open up a window of vulnerability for failures 
with byzantine outcomes. But, certain popular libraries that leverage MPI  are evidently not fault tolerant.
Nowadays, fault detection with automatic recovery without batch requeueing is a strong requirement for production environments. Thus, allowing deadlock and setting long timeouts  are suboptimal for fault detection  even when paired with conservative  recovery from the penultimate checkpoint. 

When MPI is used as a communication mechanism within a CPR library, such libraries must offer fault-tolerant extensions with minimal detection, isolation, mitigation, and potential recovery semantics to aid the CPR's library fail-backward.  Communication between MPI and the checkpoint library regarding system health may be valuable.  For fault-tolerant MPI programs (e.g.,  using APIs like FA-MPI,  Stages/Reinit, or ULFM), the checkpoint library must cooperate with the extended model or else invalidate  fault-tolerant operation.   
\end{abstract}


\section{Introduction}
Having MPI jobs killed and  re-queued at the tail of their specific job queue (using a previous checkpoint) is undesirable when it can be avoided; fault-tolerant extensions to MPI are designed, in some cases,  to avoid this situation (e.g., \cite{georgakoudis2021reinit}). This recovery mode introduces significant and unwanted delays when faults can be managed. 
But, 
if checkpoint-restart (CPR) libraries 
for MPI applications are not themselves fault tolerant, they expose MPI programs to secondary risks of byzantine failures. 
Further, there is no guarantee that fail-stop occurs. Deadline can occur, and, unless the application is actively monitored, or has a global time-out switch, it will simply waste the remainder of its scheduled allocation.

CPR libraries that use MPI need to collaborate with the MPI implementation, at least minimally, to support fail backward and avoid deadlocks when failures occur during their control of the parallel program.
They must also collaborate with the fault-tolerance model used in an MPI program augmented with a specific set of fault-tolerant extensions. 
The fault-tolerant MPI model of the CPR library has to be appropriate to that fault-tolerance model, and not be incompatible with any and all fault-tolerant actions taken by the application.  Otherwise destructive interference between the concerns will occur, and prevent them from  collaborating during detection, isolation, mitigation, and recovery. 
Fault tolerant models that use a checkpoint-restart library to checkpoint MPI state, in addition to application state, add to the importance of resilience of the checkpoint-restart library itself.
This use case may pose other  requirements too.

The remainder of this short paper is organized as follows:
Section~\ref{sec:Challenges-Opportunities} identifies  issues and concerns about fault-tolerance in a CPR library, its interactions with MPI, as well as interactions with the MPI application.  Section~\ref{sec:path-forward} 
offers straightforward next steps to advance CPR libraries.
 Conclusions and some future work are discussed in Section~\ref{sec:conclusions}.

\section{Identifying the Issues} 
\label{sec:Challenges-Opportunities}

Multiple fault-tolerant models appear to be inherent  if checkpoint-restart is employed with a fault-tolerant model (which is common) and the CPR library also utilizes MPI (which varies between different products).
Only if a given checkpoint library is MPI-free can one evidently avoid the multiple MPI fault-tolerant model issue.  

One upshot of fault tolerance within the CPR library is that, if a given MPI-dependent checkpoint library uses one specific  model only, then that model fixes what the application may itself use.  Alternatively, a given CPR library would have to be configurable for each kind of support fault-tolerant model used in a set of production applications, which would be expensive to code, validate, and maintain.
    
For instance, with MPI Stages (a Reinit-type  model) \cite{georgakoudis2021reinit,stages-1}, the approach is to quiesce MPI and then call the checkpoint library for application state preservation plus optional MPI-Stages-related serialization strings for future recovery; then, the approach is to call a separate MPI-Stages internal state checkpoint (including application objects serialized). This concept of operations apparently does not contemplate that the CPR library actually uses MPI itself, thereby changing the internal state of MPI.  Further, if a fault occurs during this checkpoint procedure, it is not guarded by a resilient MPI mode of operation that may detect faults/failures and provide potential for an isolation, mitigation, or recovery process.  Such a failure window diminishes the benefit of the MPI Stages (or other fault-detecting operational model or system). In short, the application could still crash, abort, or deadlock.

Whenever a CPR library uses MPI, it should use the fault-tolerance extensions available to that MPI implementation, or at least work within standard MPI's ability to return error codes and assess continuability after errors.
In particular,    a CPR library's 
use of MPI will need to utilize  
a) a fault detection mechanism, b) a fault-friendly means to reach consensus between peer processes in the given MPI communicator's scope (or the entire world of processes), c) a means to cause the MPI session to fail-backward, and, d) a way to
pass back error information to the application, if the application should commit to the fail-backwards itself (likely).
A CPR library's
specific fault-tolerant syntax and semantics will either have to  match  the application's 
or run in a mode where multiple, interoperable fault-tolerant libraries are present. This introduces the fault-tolerant-library concern that real applications have, even though it is part of defensive I/O to support fault tolerance itself.
Further, if the checkpoint library uses MPI I/O rather than POSIX I/O, then the entire MPI I/O feature set used has to be fault tolerant as well, which is a heavy lift in terms both of specification and implementation.  

A reasonable fault-tolerance model for a CPR library  would be as follows, leveraging  MPI's C interface,  errors-return mode, and some nominal FT extension(s) in any of the common models\footnote{  ULFM \cite{10.1007/978-3-642-33518-1_24}, Reinit \cite{georgakoudis2021reinit}, MPI Stages \cite{stages-1,stages-2}, and FA-MPI \cite{DBLP:conf/sc/HassaniSBB15} are deemed common models since discussed in the MPI Forum's Fault Tolerance Working Group.}:
\begin{itemize}
\item require that the MPI program be initialized to use \texttt{MPI\_Errors\_return} and sample all  return codes
    \item use fault-tolerant allgathers or reductions to achieve consensus on error state among surviving processes 
    \item take one of these two actions: 
    \begin{enumerate} \item return a meaningful error code (across all surviving processes) to the application so that the application realizes it must activate its own fault tolerance mode. The application has to interpret that code in terms of the fault tolerant model it is itself using.
    \item
    directly initiate recovery in the same model that the application uses\footnote{It will probably be needed to code variations into the CPR library, but it seems expensive at this time to force the library to have a new, manual coding approach for every application FT mode chosen}.
    \end{enumerate}
\end{itemize}

An alternative to the error-return code is to recompile the CPR library with C++, while still using the MPI C interface, but augmenting it with  
additional C++ features
that provide fault-tolerant extensions (e.g., exceptions)\cite{DBLP:journals/corr/abs-2107-10566}.  In this way, a group-wide parallel containment domain could be supported
and the CPR library could determine consensus in that containment domain, and then simply rethrow the exception to the application.  A common parallel containment domain methodology usable with C++ and Stages or FA-MPI appears doable at present, meaning that the CPR library would be recoded once to support either model.  ULFM as currently evolving in the MPI Forum may also be applicable.

At  least, one would need to provide the following to make the MPI usage inside a checkpoint library ``ready'' for use in conjunction with the application's fault-tolerance model:
\begin{itemize}
    \item All collective routines must be able to work in a fault-tolerant mode, so they do not deadlock when a network error occurs or a process hangs or fails\footnote{Such a mode of operation is being considered at present in the Fault Tolerant Working Group of the MPI Forum.}; 
    \item A mechanism of consensus for the health state (failed, good, unknown) must be delivered to the CPR library whenever a collective operation it initiates fails.
    \item If there should be errors in point-to-point operations, the CPR library must be able to invoke a consensus-type collective operation to determine the state of the MPI communicator (usually \texttt{MPI\_COMM\_WORLD}).
    \item If the CPR library uses MPI I/O, then the MPI I/O library must be minimally fault tolerant with regard to returning errors codes, not deadlock with dead or hung processes, and time out when I/O services become unresponsive.
\end{itemize}
In all cases, we expect that such failures detected at the CPR library level will generate one of these actions:
\begin{itemize}
    \item fail-upward back to the MPI application, so it returns from checkpoint with consensus to stop running and uses a Reinit-type decision and recovers (without re-queueing by the job scheduler) with an earlier checkpoint
    \item directly cause the application to do the Reinit-type operation as a policy of the checkpoint library
    \item kill/abort the application, and let it be restarted by the job scheduler with an earlier checkpoint. (As noted, this is the most expensive option from the user's viewpoint.)
\end{itemize}
Interactions between the CPR library and MPI implementation also appear to have merit for future consideration in failure modes. For instance, overhead reductions could be possible.


\section{Path Forward}
\label{sec:path-forward}
A key next step appears to be a careful review of how popular 
CPR libraries
use MPI
in detail
as a first step to adding fault-tolerant extensions to them. 
The choice of fault-tolerant modes of these libraries will either need to reflect the choice of a single overarching fault-tolerant mode for MPI, or else support multiple ones.  
Prototyping fault-tolerant extensions within a CPR library like SCR \cite{LLNLSCR}  appears warranted at present, with demonstration of overall resilience when coupled with a fault-tolerant MPI program.

What has noted been noted so far is that these libraries use MPI, but do nothing overt to change error-handlers, sample MPI API return codes, or exploit fault-tolerant extensions: SCR \cite{LLNLSCR} and MANA \cite{garg2019mana}.
We noted that BLCR and DMTCP \cite{hargrove2006berkeley,DBLP:journals/corr/abs-cs-0701037} do not use MPI directly.
Without having yet evaluated its  performance or examined its implementation, 
the meta-CPR library 
CRAFT \cite{CRAFT} describes using SCR and ULFM together for fault-tolerant operation, 
and is exemplary of what this paper proposes as a baseline requirement set/approach  for CPR libraries.

Transparent CPR libraries (e.g., DMTCP \cite{DBLP:journals/corr/abs-cs-0701037} and MANA
\cite{garg2019mana}) may use MPI in ways that introduce deadlock when an MPI operation fails.  Careful study of how transparent checkpointing interacts with a fault-tolerant model of MPI is also warranted. Two-way communication between a fault-tolerant MPI and a transparent CPR library appears useful for reduced overheads too.  

Lastly, CPR libraries must minimally become aware of new error information and the pre-\texttt{MPI\_Init} ability to set error handlers supported by MPI-4, as well as MPI Sessions. The paper by Skjellum et al., considers fault-tolerance and Sessions \cite{stages-3}. This paper also suggests an extended execution model for fault-tolerant MPI programs; that explicit scaffolding   could also be useful for exploitation by CPR libraries as well, particularly to help automate unsupervised fallback recovery without batch requeueuing.   

\section{Conclusion}
\label{sec:conclusions}
This short paper discussed and motivated the need for checkpoint-restart (CPR) libraries to become more fault tolerant when used in conjunction with MPI applications and libraries.  Further requirements are posed on the library when the MPI program is extended with fault-tolerant extensions.

It is well known that 
 checkpoint-restart libraries  must  be fault tolerant, not just deadlock or timeout, or else they open up a window of vulnerability for failures to occur with byzantine outcomes. But, certain popular libraries that leverage MPI  are evidently not fault tolerant. 
Nowadays, fault detection with automatic recovery without batch requeueing form a strong requirement for production environments. Thus, allowing deadlock and relying on  long timeouts for fault detection  is a  suboptimal strategy   even when paired with conservative, but common  recovery policy of using the penultimate checkpoint, rather than a the last checkpoint (which might have been compromised by a fault during creation).

When MPI is used as a communication mechanism within a CPR library, such libraries must integrate fault-tolerant extensions with minimal detection, isolation, mitigation, and potential recovery semantics to aid the CPR's library fail-backward.  Communication between MPI and the checkpoint library regarding system health may be valuable.  For fault-tolerant MPI programs (e.g.,  using APIs like FA-MPI,  Stages/Reinit, or ULFM), the checkpoint library must cooperate with the extended model or else invalidate  fault-tolerant operation, even if they don't use MPI directly at present.

Carefully evaluating how CRAFT \cite{CRAFT} works with and integrates SCR \cite{LLNLSCR} and ULFM \cite{10.1007/978-3-642-33518-1_24}, then extrapolating how it could work with other fault tolerant models of MPI, will be among the future work to be explored. Refining the set of  interoperability requirements and data interchanges needed for Reinit/Stages and FA-MPI to work well with a CPR library is also planned as extensions to the principles described more generally here.  Determining whether CPR libraries  pose new,  implicit and/or previously unfulfilled requirements on fault-tolerant models of MPI will also be considered.

\section*{Acknowledgments}
The authors wish to thank the anonymous reviewers, whose feedback helped improve the final document.

This work was performed with partial support from the National Science
Foundation  under Grants Nos.~1822191, 1918987, 1925538, and 1925603. Any opinions, findings, and conclusions or recommendations expressed in this material are those of the authors and do not necessarily reflect the views of the National Science Foundation.

\nocite{osti_1314607}
\bibliographystyle{IEEEtran}
\bibliography{references}

\end{document}